\documentclass[twocolumn,showpacs,superscriptaddress,amsmath,amssymb,nofootinbib,aps, prd]{revtex4-2}
\usepackage{graphicx}
\usepackage{epsfig}
\usepackage{dcolumn}
\usepackage{multirow}
\usepackage{booktabs}
\usepackage{appendix}
\usepackage{bm}
\usepackage{overpic}
\usepackage{color}
\usepackage{subfigure}
\usepackage{threeparttable}
\usepackage{lineno}
\usepackage[colorlinks,linkcolor=blue,anchorcolor=blue,citecolor=blue]{hyperref}
\usepackage{epstopdf}

\def\pip{\pi^{+}}
\def\pim{\pi^{-}}
\def\piz{\pi^{0}}
\def\ks{K_{S}^{0}}
\def\ee{e^{+}e^{-}}
\def\dedx{\mathrm{d}E/\mathrm{d}x}

\def\BR{\mathcal{B}}

\def \gev  {\mbox{GeV}}
\def \gevc {\mbox{GeV/$c$}}
\def \gevcc{\mbox{GeV/$c^2$}}
\def \mev  {\mbox{MeV}}

\def \ifb  {\mbox{fb$^{-1}$}}

\def \mbc {M_{\rm{BC}}}
\def \msigmap {M_{p\pi^0}}

\def \dE {\Delta E}

\def \ebeam {E_{\rm{beam}}}

\def \romanOne   {\uppercase\expandafter{\romannumeral1}}
\def \romanTwo   {\uppercase\expandafter{\romannumeral2}}
\def \romanThree {\uppercase\expandafter{\romannumeral3}}
\def \romanFour  {\uppercase\expandafter{\romannumeral4}}
\def \romanFive  {\uppercase\expandafter{\romannumeral5}}
\def \romanSix   {\uppercase\expandafter{\romannumeral6}}
\def \romanSeven {\uppercase\expandafter{\romannumeral7}}
\def \romanEight {\uppercase\expandafter{\romannumeral8}}
\def \romanNine {\uppercase\expandafter{\romannumeral9}}

\newcommand{\lamcplamcm}{\Lambda_{c}^{+}\bar{\Lambda}_{c}^{-}}
\newcommand{\lambdacp}{\Lambda_{c}^{+}}
\newcommand{\lambdacm}{\bar{\Lambda}_{c}^{-}}
\def\kshort{K^0_{\mathrm{S}}}

\newcommand{\sigmode}[1]{
	\ifnum#1=1
	\lambdacp \rightarrow \Sigma^0 K^+
	\else
	\ifnum#1=2
	\lambdacp \rightarrow \Sigma^+ \kshort
	\fi
	\fi
}
\newcommand{\refmode}[1]{
	\ifnum#1=1
	\lambdacp \rightarrow \Sigma^0 \pi^+
	\else
	\ifnum#1=2
	\lambdacp \rightarrow \Sigma^+ \pi^+\pi^-
	\fi
	\fi
}
\newcommand{\sigmodefs}[1]{
	\ifnum#1=1
	\Sigma^0 K^+
	\else
	\ifnum#1=2
	\Sigma^+ \kshort
	\fi
	\fi
}
\newcommand{\refmodefs}[1]{
	\ifnum#1=1
	\Sigma^0 \pi^+
	\else
	\ifnum#1=2
	\Sigma^+ \pi^+\pi^-
	\fi
	\fi
}

\newcommand{\bkgmode}{
	\lambdacp \rightarrow p \ks \piz
}

\begin{document}
\title{\bf\boldmath Measurement of Branching Fractions of Singly Cabibbo-suppressed Decays $\Lambda_c^+ \rightarrow \Sigma^{0} K^+$ and $\Sigma^{+} K_{S}^0$}
\author{\small
M.~Ablikim$^{1}$, M.~N.~Achasov$^{11,b}$, P.~Adlarson$^{70}$, M.~Albrecht$^{4}$, R.~Aliberti$^{31}$, A.~Amoroso$^{69A,69C}$, M.~R.~An$^{35}$, Q.~An$^{66,53}$, X.~H.~Bai$^{61}$, Y.~Bai$^{52}$
% ...
\\
\vspace{0.2cm}
(BESIII Collaboration)\\
\vspace{0.2cm} {\it
$^{1}$ Institute of High Energy Physics, Beijing 100049, People's Republic of China\\
% ...
$^{4}$ Bochum Ruhr-University, D-44780 Bochum, Germany\\
% ...
$^{11}$ G.I. Budker Institute of Nuclear Physics SB RAS (BINP), Novosibirsk 630090, Russia\\
% ...
$^{31}$ Johannes Gutenberg University of Mainz, Johann-Joachim-Becher-Weg 45, D-55099 Mainz, Germany\\
% ...
$^{35}$ Liaoning Normal University, Dalian 116029, People's Republic of China\\
% ...
$^{52}$ Southeast University, Nanjing 211100, People's Republic of China\\
$^{53}$ State Key Laboratory of Particle Detection and Electronics, Beijing 100049, Hefei 230026, People's Republic of China\\
% ...
$^{61}$ University of Jinan, Jinan 250022, People's Republic of China\\
% ...
$^{66}$ University of Science and Technology of China, Hefei 230026, People's Republic of China\\
% ...
$^{69}$ University of Turin and INFN, (A)University of Turin, I-10125, Turin, Italy; (B)University of Eastern Piedmont, I-15121, Alessandria, Italy; (C)INFN, I-10125, Turin, Italy\\
$^{70}$ Uppsala University, Box 516, SE-75120 Uppsala, Sweden\\
% ...
$^{b}$ Also at the Novosibirsk State University, Novosibirsk, 630090, Russia\\
% ...
}\vspace{0.4cm}}
\date{\today}
\begin{abstract}
Based on a sample of 4.4 $\mathrm{fb}^{-1}$ of $e^{+}e^{-}$ annihilation data collected in the energy region between 4.6 GeV and 4.7 GeV with the BESIII detector at BEPCII, two singly Cabibbo-suppressed decays $\Lambda_c^+ \rightarrow \Sigma^0 K^+$ and $\Lambda_{c}^{+} \rightarrow \Sigma^+ K_{S}^0$ are studied. 
The ratio of the branching fraction $\mathcal{B}(\Lambda_c^+ \rightarrow \Sigma^0 K^+)$ relative to $\mathcal{B}(\Lambda_c^+ \rightarrow \Sigma^0 \pi^+)$ is measured to be $0.0361 \pm 0.0073(\mathrm{stat.}) \pm 0.0005(\mathrm{syst.})$, and the ratio of $\mathcal{B}(\Lambda_c^+ \rightarrow \Sigma^+ K_{S}^0)$  relative to $\mathcal{B}(\Lambda_{c}^{+} \rightarrow \Sigma^+ \pi^+ \pi^-)$ is measured to be $0.0106 \pm 0.0031(\mathrm{stat.}) \pm 0.0004(\mathrm{syst.})$. After taking the world-average branching fractions of the reference decay channels, the branching fractions
$\mathcal{B}(\Lambda_c^+ \rightarrow \Sigma^0 K^+)$ 
and
$\mathcal{B}(\Lambda_{c}^{+} \rightarrow \Sigma^+ K_{S}^0)$ 
are determined to be 
$(4.7\pm 0.9(\mathrm{stat.})\pm 0.1(\mathrm{syst.}) \pm 0.3(\mathrm{ref.}))\times10^{-4}$
and
$(4.8\pm 1.4(\mathrm{stat.})\pm 0.2(\mathrm{syst.}) \pm 0.3(\mathrm{ref.}))\times10^{-4}$, respectively.
The branching fraction of the $\Lambda_{c}^{+} \rightarrow \Sigma^+ K_{S}^0$ decay is measured for the first time.
\end{abstract}
\maketitle

\setrunninglinenumbers
	
\section{\boldmath Introduction}
The study of charmed baryon decays is valuable for both understanding charmed-baryon dynamics and probing the effects of the weak and strong interactions.
Since the ground state of the charmed baryons $\Lambda_{c}^+$ was discovered~\cite{Abrams:1979iu}, many efforts have been made to predict its branching fractions (BFs) into two-body hadronic final states,  $\Lambda_{c}^+ \rightarrow \mathrm{B}_{n} M$~\cite{Uppal:1994pt, Zou:2020, Zhao:2020, Geng:2019xbo,he2021global,Cheng:2021qpd}, where $\mathrm{B}_{n}$ and $M$ denote the octet baryon and nonet meson states, respectively. However, progress  has been hindered, due to the  limited precision of experimental measurements~\cite{al_2006} and difficulties in the theoretical treatment of non-perturbative strong interaction effects. For example, before studies of Singly Cabibbo-Suppressed (SCS) $\Lambda_{c}^+$ decays were performed by the Belle~\cite{Abe:2001mb} and BaBar~\cite{Aubert:2006wm} collaborations,  only one theoretical calculation existed for the BFs of these SCS processes~\cite{Uppal:1994pt}.
Throughout this paper, the charge-conjugate modes are implied, unless otherwise stated.

The challenge for theoretical predictions is that the well-known factorization approach, which has been applied successfully to heavy-meson decays, has difficulties in describing charmed-baryon decays~\cite{Cheng:2018}. This is because the nonfactorizable terms are sizable or even dominant contributions in $\Lambda_{c}^+$ hadronic decays~\cite{Uppal:1994pt}, compared with the charmed-meson case~\cite{Li:2012cfa,Saur:2020rgd}, \emph{e.g.}, the SCS $\sigmode{1}$ and $\sigmodefs{2}$ decays {\it only} receive nonfactorizable contribution. These nonfactorizable terms, arising from $W$-exchange or internal $W$-emission~\cite{Zou:2020}, can be constrained by precise experimental inputs in charmed-meson decays~\cite{Bianco:2003vb}.

The BESIII collaboration~\cite{Ablikim:2016} has reported significant improvements in the precision of absolute hadronic BFs of the $\Lambda_{c}$ baryon, and the first model-independent measurements near the threshold of $\lamcplamcm$ production~\cite{Li:2021iwf}. In addition, the BESIII, LHCb and Belle collaborations have carried out complementary analyses of charmed baryons, such as lifetime measurements~\cite{LHCb:2018nfa,LHCb:2019ldj,LHCb:2021vll,Cheng:2021vca,lifetime-BelleII} and studies of semi-leptonic decays~\cite{BESIII:2015ysy,BESIII:2016ffj,BESIII:2018mug,Belle:2021crz,Belle:2021dgc}.
Improved measurements of the BFs of the $\Lambda_{c}$ decays can provide crucial inputs for the theoretical models~\cite{Zou:2020, Zhao:2020, Geng:2019xbo,he2021global}, in particular those of the SCS $\Lambda_{c}\to\Sigma K$ decays for which there exists very limited experimental information.

Theoretical predictions for SCS $\Lambda_c$ decays are listed in Table~\ref{tab:k0 prediction}.
In Refs.~\cite{Uppal:1994pt, Zou:2020}, factorizable terms are made accessible by inserting vacuum intermediates states~\cite{Uppal:1994pt}, which are then reduced to the products of current matrix elements defined with decay constants of the emitted meson $M$ and form factors of the $\Lambda_{c}^+\to\mathrm{B}_{n}$ transition. Nonfactorizable terms are tackled in the current algebra framework with the pole approximation in Refs.~\cite{Uppal:1994pt, Zou:2020}. Ref.~\cite{Zou:2020} uses the MIT bag model~\cite{MIT:1974} to account for baryon pole transition matrix elements, while Ref.~\cite{Uppal:1994pt} makes short-distance QCD corrections to the weak Hamiltonian. 
In Ref.~\cite{Zhao:2020}, a diagrammatic analysis is performed and $\BR(\sigmode{2})$ is predicted to be $(9.6\pm2.4)\times10^{-4}$. Ref.~\cite{Geng:2019xbo} expects that $\BR(\sigmode{1}) = \BR(\sigmode{2}) = (5.4\pm0.7)\times10^{-4}$ under $SU(3)_F$ flavor symmetry. In Ref.~\cite{he2021global},  the irreducible representation amplitude (IRA) approach is used to extract amplitudes from experimental data inputs, which gives quite different predictions for $\BR(\sigmode{1})$ and $\BR(\sigmode{2})$.  In particular, Ref.~\cite{he2021global} 
predicts that $\BR(\sigmode{2})$ is about one fifth of $\BR(\sigmode{1})$, which is far smaller than other predictions.

Table~\ref{tab:k0 prediction} shows the current Particle Data Group (PDG)~\cite{Zyla:2020zbs} world average value of $\BR(\sigmode{1})$ based on measurements from the Belle~\cite{Abe:2001mb} and BaBar~\cite{Aubert:2006wm} collaborations, performed more than a decade ago, while no measurement exits for the $\sigmode{2}$ decay.   All theoretical predictions for $\BR(\sigmode{1})$ are consistent with the experimental value.   Those predictions in Refs.~\cite{Zhao:2020, Geng:2019xbo,he2021global} are from fits that take an ensemble of measured BFs as inputs, and are limited by the precision of these measurements.  Thus, new determinations of the BFs of $\Lambda_{c}\to\Sigma K$ decays, in particular the mode $\sigmode{2}$, are important for validating and improving these theoretical-model calculations. Furthermore, improved measurements may clarify the tension between the predictions in Ref.~\cite{he2021global} and Refs.~\cite{Uppal:1994pt, Zou:2020, Zhao:2020, Geng:2019xbo}.

\begin{table}[htbp]
	\caption{Comparison of various theoretical predictions and the experimental values for $\BR(\lambdacp\to\Sigma K)$ (in unit of $10^{-4}$). In Ref.~\cite{Uppal:1994pt}, alternative assignments to QCD corrections give different predictions as shown in the parentheses. The theoretical uncertainties in Ref.~\cite{Zou:2020} are estimated to be 25\%, arising from a slight change of the MIT bag radius.  \label{tab:k0 prediction}}
	\setlength\tabcolsep{0.1pt}
	\begin{center}
		\begin{tabular}{cc|c}
			\hline\hline
			 & $\BR(\sigmode{1})$ &  $\BR(\sigmode{2})$ \\
			\hline
			QCD corrections~\cite{Uppal:1994pt} & 2(8) & 2(4) \\
			MIT bag model~\cite{Zou:2020} & $7.2\pm1.8$ & $7.2\pm1.8$ \\
			Diagrammatic analysis~\cite{Zhao:2020} & $5.5\pm1.6$ & $9.6\pm2.4$ \\
			$SU(3)_F$ flavor symmetry~\cite{Geng:2019xbo} & $5.4\pm0.7$ & $5.4\pm0.7$ \\
			IRA method~\cite{he2021global} & $5.0\pm0.6$ & $1.0\pm0.4$ \\
			PDG 2020~\cite{Zyla:2020zbs} & $5.2\pm0.8$ &   / \\
			\hline\hline
		\end{tabular}
	\end{center}
\end{table}

In this paper, we present a study of the SCS decays $\sigmode{1}$ and $\sigmodefs{2}$ based on 4.4\,$\ifb$ of $\ee$ annihilation data collected at the center-of-mass energies  $\sqrt{s}=4.600$, 4.612, 4.628, 4.641, 4.661, 4.682, 4.699 $\gev$~\cite{BESIII:2015qfd,BESIII:2022ulv} with the BESIII detector at BEPCII. We report the first study of the channel $\sigmode{2}$ and provide the BF ratio, $\mathcal{B}(\sigmode{2})/\mathcal{B}(\refmode{2})$, together with an improved measurement of the BF ratio, $\mathcal{B}(\sigmode{1})/\mathcal{B}(\refmode{1})$.


\section{\boldmath BESIII Experiment and Monte Carlo Simulation}
The BESIII detector~\cite{Ablikim:2009aa} records symmetric $e^+e^-$ collisions provided by the BEPCII storage ring~\cite{Yu:2016cof} in the center-of-mass energy range from 2.0 to 4.95~GeV, with a peak luminosity of $1 \times10^{33}\;\text{cm}^{-2}\text{s}^{-1}$ achieved at $\sqrt{s} = 3.77\;\text{GeV}$. The cylindrical core of the BESIII detector covers 93\% of the full solid angle and consists of a helium-based multilayer drift chamber~(MDC), a plastic scintillator time-of-flight system (TOF), and a CsI(Tl) electromagnetic calorimeter (EMC), which are all enclosed in a superconducting solenoidal magnet providing a 1.0 T magnetic field. The solenoid is supported by an octagonal flux-return yoke with resistive plate counter muon identification modules interleaved with steel. The charged-particle momentum resolution at $1\,\gevc$ is $0.5\%$, and the ionization energy loss $\dedx$ resolution is $6\%$ for electrons from Bhabha scattering. The EMC measures photon energies with a resolution of $2.5\%$ ($5\%$) at $1\,\gev$ in the barrel (end-cap) region. The time resolution in the TOF barrel region is 68\,ps, while that in the end-cap region is 110\,ps. The end-cap TOF system was upgraded in 2015 using multi-gap resistive plate chamber technology, providing a time resolution of 60\,ps~\cite{li:TOF1,*guo:TOF2,*Cao:2020ibk}. More detailed descriptions can be found in Refs.~\cite{detector,BEPCII}.

Simulated data samples are produced with a {\sc Geant4}-based~\cite{geant4} Monte Carlo (MC) package, which includes the geometric description of the BESIII detector~\cite{GDMLMethod,BesGDML} and the detector response. The simulation models the beam-energy spread and initial-state radiation (ISR) in the $\ee$ annihilations with the generator {\sc kkmc}~\cite{kkmc}. The final-state radiation from charged final-state particles is incorporated using {\sc photos}~\cite{photos}.

The ``inclusive MC sample'' includes the production of $\lamcplamcm$ pairs and  open-charmed mesons, ISR production of vector charmonium(-like) states, and continuum processes which are incorporated in {\sc kkmc}~\cite{kkmc, kkmc2}. Known decay modes are modeled with {\sc evtgen}~\cite{evtgen, besevtgen} using the BFs taken from the PDG~\cite{Zyla:2020zbs}. The BF $\mathcal{B}(\sigmode{2})$ is assumed to be the same as $\mathcal{B}(\sigmode{1})$. The remaining unknown charmonium decays are modeled with {\sc lundcharm}~\cite{lundcharm, Yang:2014vra}. The inclusive MC sample is used to study background contributions and to optimize event selections. We denote ``Hadron MC'' as the inclusive MC sample with $\lamcplamcm$ pairs removed, which therefore only includes backgrounds for this study.
For the reference mode $\Lambda_{c}^{+} \rightarrow \Sigma^{+} \pi^{+} \pi^{-}$, the intermediate states are modeled according to an internal partial-wave analysis of this channel. For the reference mode $\Lambda_{c}^{+} \rightarrow \Sigma^{0} \pi^{+}$, the angular distributions are described with consideration of the transverse polarization and decay asymmetry parameters of the  $\Lambda_{c}^{+}$ and its daughter baryons~\cite{Ronggang:Lambda_c}. We use a uniformly distributed phase-space model for the simulation of the signal SCS decays $\Lambda_{c}^+\to\Sigma^0K^+$ and  $\Sigma^+K_{S}^0$. The $\ee\to\lamcplamcm$ ``signal MC'' samples, in which the $\Lambda_{c}^{+}$ decays exclusively into signal (reference) modes while the $\bar{\Lambda}_{c}^{-}$ decays inclusively, are used to determine the detection efficiencies.

\section{\boldmath Event Selection}
\label{sec:selection}
In this analysis, we reconstruct the two signal modes through the cascade decays
$\Lambda_c^+ \rightarrow \Sigma^0 K^+, \Sigma^0 \rightarrow \gamma \Lambda, \Lambda \rightarrow p \pi^-$ and $\Lambda_c^+ \rightarrow \Sigma^+ K_{S}^0, \Sigma^+ \rightarrow p \pi^0, \pi^0 \rightarrow \gamma \gamma, K_{S}^0 \rightarrow \pi^+ \pi^-$. The reference modes $\refmode{1}$ and $\Sigma^{+}\pi^+\pi^-$ are reconstructed through the same decay chains of the $\Sigma^0$ and $\Sigma^+$ baryons.  
As the $\lamcplamcm$ pair is produced without any accompanying hadrons, it is possible to reconstruct the $\lambdacp$  and infer the presence of the  $\lambdacm$ through its recoil mass.  

Charged tracks are reconstructed in the MDC, and are required to have a polar angle $\theta$ with respect to the $z$-axis, defined as the symmetry axis of the MDC, satisfying $|\!\cos\theta|<0.93$.
Those tracks that are not used in the reconstruction of $\ks\to\pip\pim$ and $\Lambda\to p\pi^-$ candidates must 
have a distance of closest approach to the interaction point (IP) smaller than 10 cm along the $z$-axis ($V_z$) and smaller than 1 cm in the perpendicular plane ($V_r$).  Particle identification (PID) for charged tracks is implemented~\cite{asner_physics_2008} using combined information from the flight time measured in the TOF and the $\dedx$ measured in the MDC. Charged tracks are identified as protons when they satisfy $\mathcal{L}(p)> \mathcal{L}(K)$, $\mathcal{L}(p)> \mathcal{L}(\pi)$ and $\mathcal{L}(p)>0.0001$, where $\mathcal{L}(h)$ is the PID probability for each hadron $(h)$ hypothesis with $h=p,\pi,K$. Charged tracks are identified as pions when they satisfy $\mathcal{L}(\pi) > \mathcal{L}(K)$. In the selection of $\sigmode{1}$ and $\refmodefs{1}$ decays, no PID requirement is imposed for the bachelor kaons and pions, but the $\chi^2_{\mathrm{PID},K(\pi)}$ for kaon (pion) hypothesis is retained for further analysis in the signal-candidate selection.

Photon candidates from $\piz$ and $\Sigma^{0}$ decays are reconstructed from the electromagnetic showers detected in the EMC crystals. The deposited energy is required to be larger than $25\,\mev$ in the barrel region with $|\!\cos\theta|<0.80$ and larger than $50\,\mev$ in the end-cap region with $0.86<|\!\cos\theta|<0.92$. To further suppress fake photon candidates due to electronic noise or beam-related background, the measured EMC time is required to be within 700 ns from the event start time. To reconstruct $\piz$ candidates, the invariant mass of a photon pair is required to satisfy $0.115<M_{\gamma\gamma}<0.150\,\gevcc$. To further improve the momentum resolution, the invariant mass of the photon pair is constrained to the 
known $\piz$ mass~\cite{Zyla:2020zbs} by applying a one-constraint kinematic fit, the $\chi^2$ of which is required to be less than 200. The momentum of the $\piz$ after the fit is used in the subsequent analysis.

Neutral kaon candidates are reconstructed through the decay $\ks\to\pip\pim$ by combining all pairs of oppositely charged tracks with both pions passing the PID requirement. These tracks, and also those used to build $\Lambda$ candidates, must satisfy $|\!\cos\theta|<0.93$ and $|V_z|<20\,\mathrm{cm}$, while no $V_r$ requirement is applied. A vertex fit is applied to pairs of charged tracks, constraining them to originate from a common decay vertex, and the $\chi^2$ of this vertex fit is required to be less than 100. The invariant mass of the $\pip\pim$ pair needs to satisfy $0.487<M_{\pip\pim}<0.511\,\gevcc$.  Here, $M_{\pip\pim}$ is calculated with the pions constrained to originate at the decay vertex. To further suppress background, we require the ratio of the decay length to the resolution of decay length to be greater than 2. An analogous normalised decay-length requirement is imposed on $\Lambda$ candidates reconstructed in the final state $p\pim$.  Here a PID requirement is imposed on the proton candidate, but not on the $\pim$ candidate.  The invariant mass of the $p\pim$ combination must satisfy $1.111<M_{p\pi^-}<1.121\,\gevcc$.

Protons and reconstructed $\piz$ mesons are used to form $\Sigma^+$ candidates.  The invariant mass of the $p\piz$ pair is required to be $1.15<M_{p\piz}<1.28\,\gevcc$, which is a loose requirement since this kinematic variable is fitted to determine the  $\Sigma^+$ signal.

When selecting $\refmode{2}$ decays we veto events where the invariant mass of the $\pi^+\pi^-$ pair satisfies $0.48<M_{\pip\pim}<0.52\,\gevcc$,  in order to avoid cross-contamination between   $\sigmode{2}$ and $\refmode{2}$.
Possible backgrounds from $\Lambda\to p\pi^{-}$ in the final states are also rejected by requiring $M(p\pi^-)$ lies outside the range $(1.11, 1.12)\,\gevcc$. Since the $\Lambda$ background is not significant in $\sigmode{2}$  decays, a $\Lambda$ veto is not imposed for the mode. If there are multiple $\Lambda_{c}^+$ combinations in a single event, we choose the candidate with the minimum magnitude of the energy difference, defined as 
$\dE \equiv E_{\Lambda_c}-\ebeam $,
where $E_{\Lambda_c}$ is the energy of the detected $\lambdacp$ candidate in the rest frame of the initial $\ee$ collision system, and $\ebeam$ is the beam energy. Furthermore, the requirement $-0.02<\dE<0.01\,\gev$ is imposed as illustrated in Fig.~\ref{fig:deltaE}, where the cut values are decided from inspection of the  the figure-of-merit $\mathrm{FOM}=\frac{S}{\sqrt{S+B}}$, where $S$ $(B)$ is the number of  signal (background) events in the inclusive MC samples.

The $\Sigma^0$ candidates are reconstructed from photon candidates and $\Lambda$ candidates. To suppress backgrounds from energetic showers, we further require the deposited energies of the photon candidates to be less than $0.25\,\gev$ in $\Sigma^{0}$ reconstruction. The invariant mass of the $\gamma\Lambda$ pair is required to lie within $1.179<M_{\gamma\Lambda}<1.203\,\gevcc$. 


When reconstructing the $\sigmode{1}$ and $\refmodefs{1}$ signal decays, we perform a kinematic fit that constrains the invariant mass of the recoil side of the $\lambdacp$ candidate to the known mass of the $\lambdacm$~\cite{Zyla:2020zbs}. To improve the resolution of photon momenta originating from $\Sigma^{0}$, we also constrain the invariant masses of the $p\pi^-$ and  $\gamma p\pi^-$ to the known masses of $\Lambda$ and  $\Sigma^0$, respectively. The fitted momenta from this three-constraint (3C) kinematic fit are used in the subsequent analysis. 
The aforementioned $\dE$ metric can be used for selection the only candidate when there are multiple $\Lambda_{c}^+$ candidates in the event. But the $\chi^2_{\mathrm{total}}  \equiv \chi^2_\mathrm{3C} + \chi^2_\mathrm{VF}  + \chi^2_{\mathrm{PID},h} $ metric gives better signal significance, where $\chi^2_\mathrm{3C}$ is the 3C fit quality and, $\chi^2_\mathrm{VF}$ is the quality of the vertex fit of the $\Lambda$ reconstruction. The $\chi^2_{\mathrm{PID},h}$ is defined as~\cite{asner_physics_2008}
\begin{equation}
    \chi^2_{\mathrm{PID},h} \equiv \left(\frac{t-t_{h}}{\sigma_{\mathrm{TOF}}}\right)^{2} + \left(\frac{\dedx-(\dedx)_{h}}{\sigma_{\dedx}}\right)^{2} \, ,
\end{equation}
where $t$ and $\dedx$ are the measured TOF and ionization 
 energy loss, the index ``$h$'' denotes the values
 expected for the $h=\pi,K$ hypotheses and $\sigma_{\mathrm{TOF}}$ and 
 $\sigma_{\dedx}$ are the resolutions of TOF and $\dedx$ measurements, respectively. No explicit PID requirement is imposed on the particle that 
  accompanies the $\Sigma^0$, and the assignment of $\sigmode{1}$ and 
  $\refmodefs{1}$ modes is only based on $\chi^2_\mathrm{total}$ with the $K$ and $\pi$ hypotheses, respectively. 
To suppress backgrounds, a requirement of $\chi^2_{\mathrm{total}}<20$  is imposed for both modes, which gives best FOM values.

\begin{figure*}[htbp]\centering
	\includegraphics[width=0.4\textwidth]{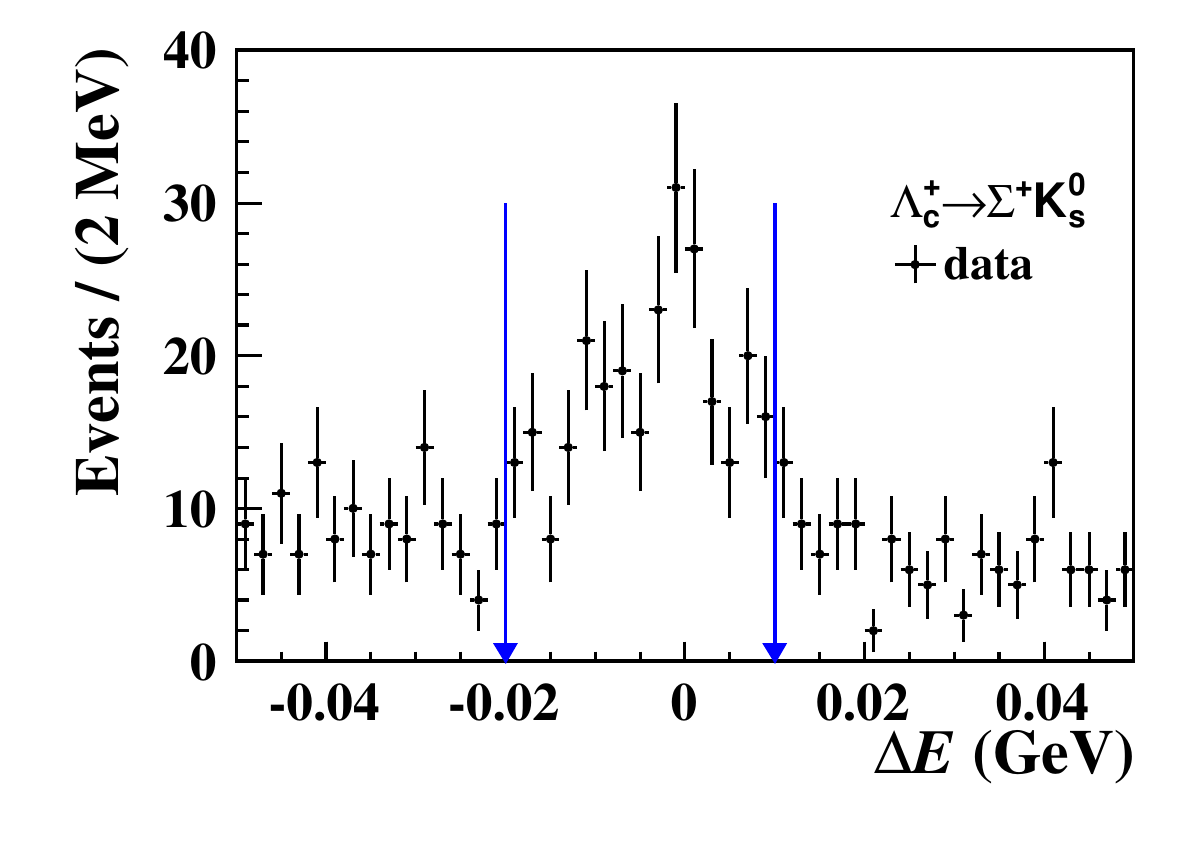}
	\includegraphics[width=0.4\textwidth]{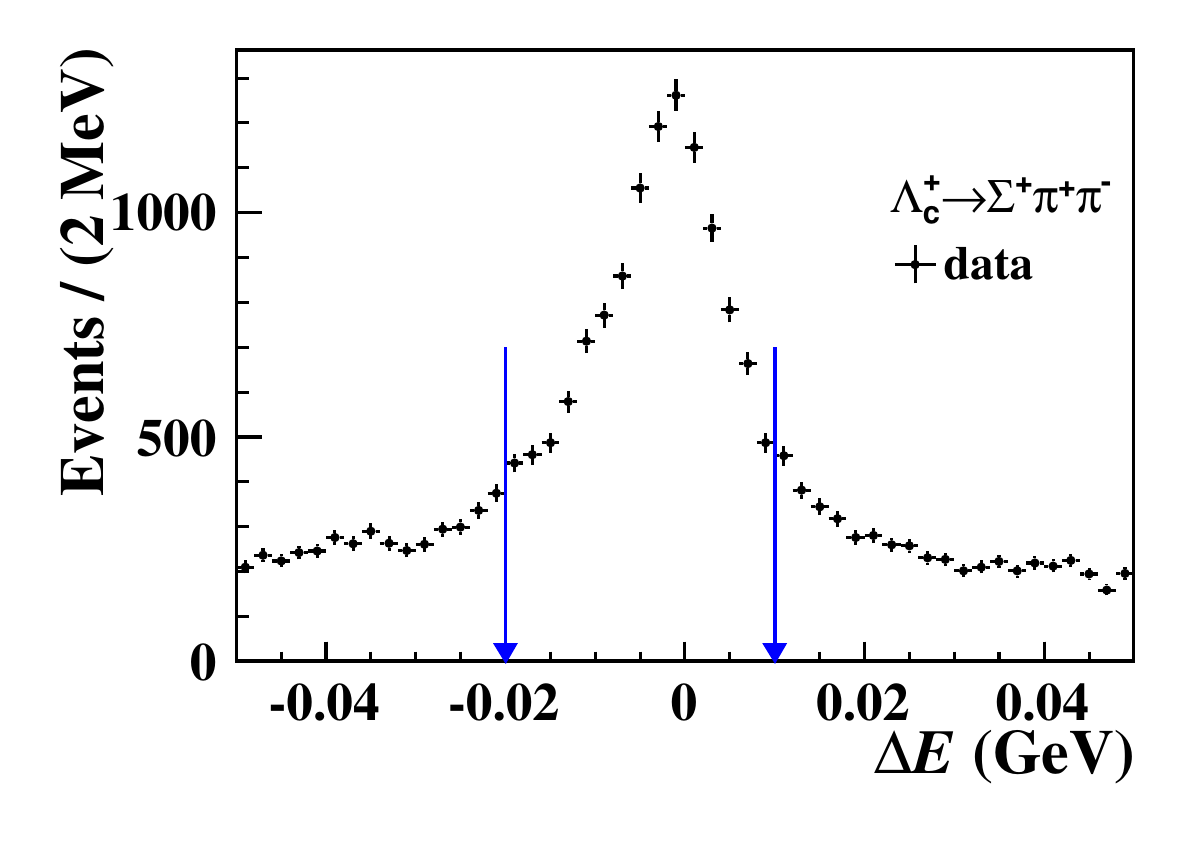}
	
	\includegraphics[width=0.4\textwidth]{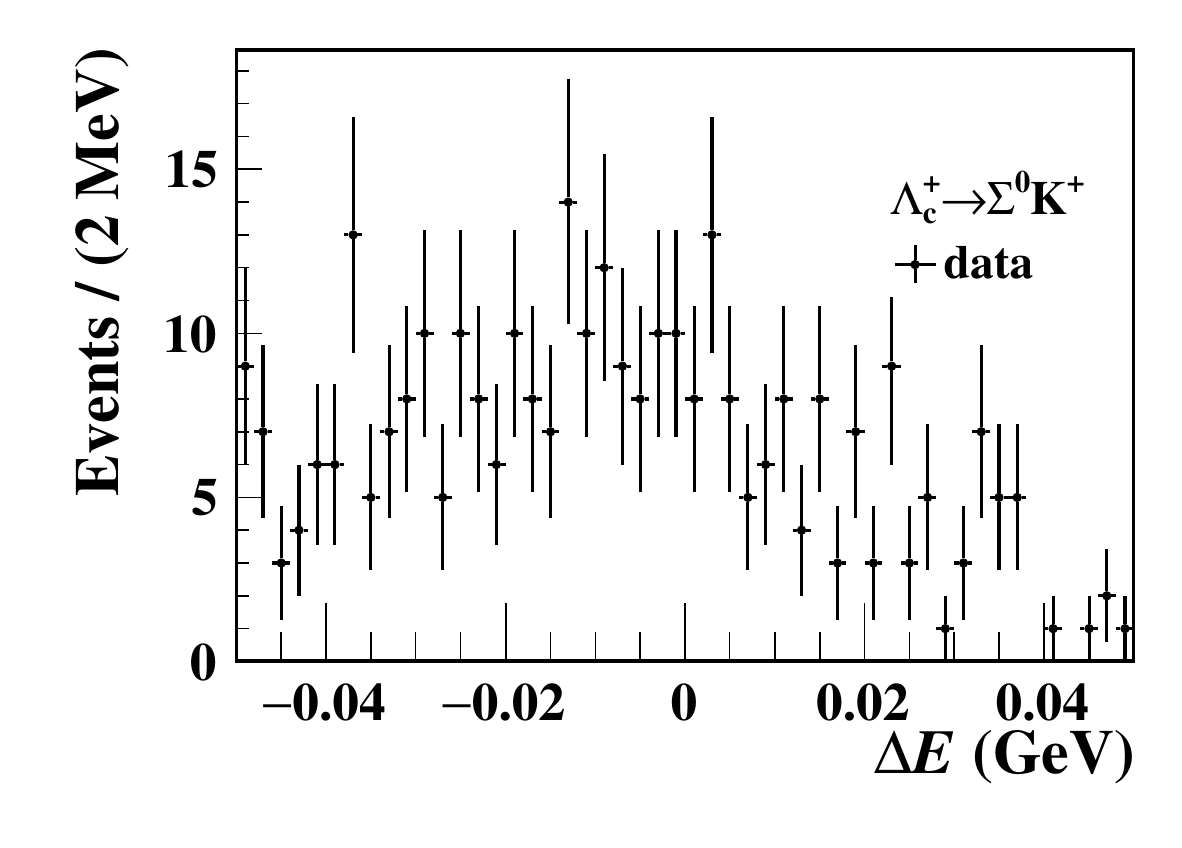}
	\includegraphics[width=0.4\textwidth]{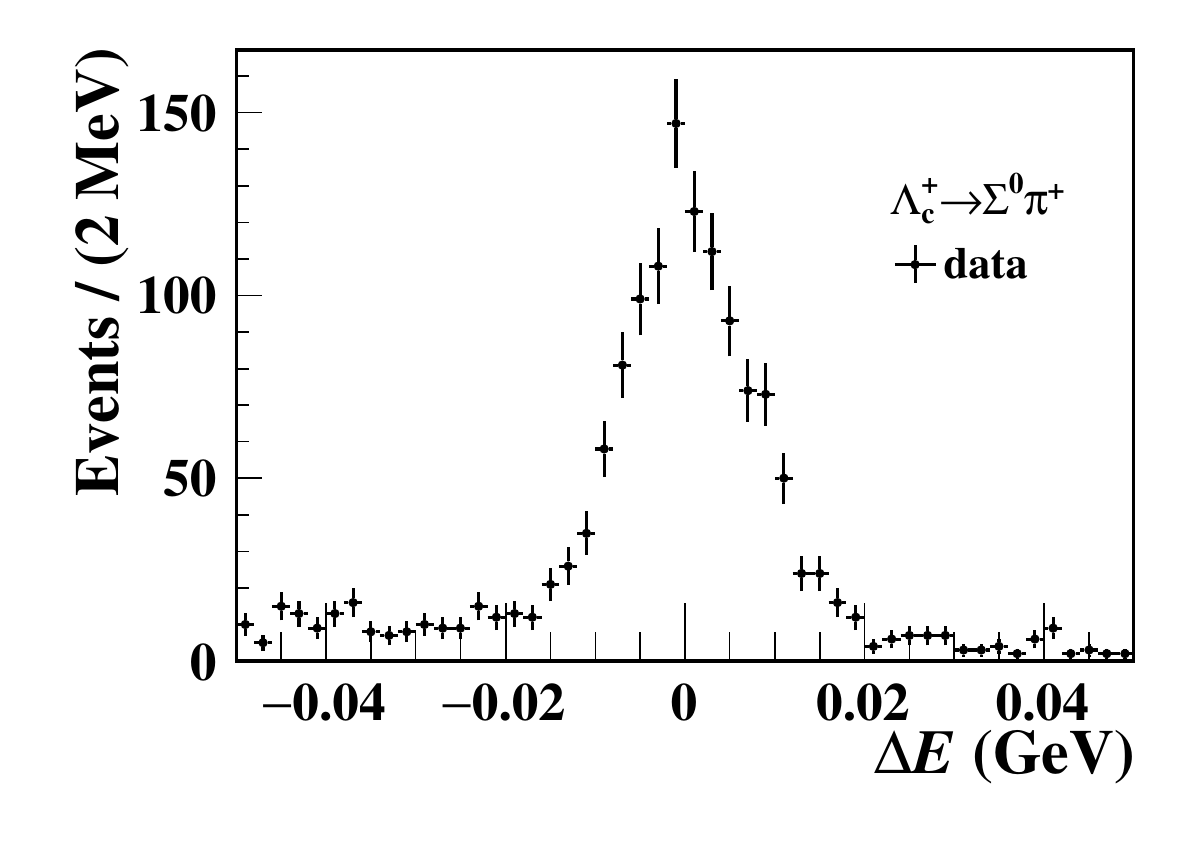}
	\caption{$\dE$ distributions for the data samples passing the event selections. The arrows indicate the requirements for $\dE$. $\mbc$ and $M_{p\pi^0}$ for $\sigmode{2}$ and $\refmodefs{2}$ modes are required to be within $\pm3\sigma$ resolutions around individual peaks. The momenta before the kinematic fit are used for calculating $\dE$ of $\sigmode{1}$ and $\refmodefs{1}$ modes.\label{fig:deltaE}}
\end{figure*}

\section{Relative branching fraction measurements}


To determine the yields for the four signal $\lambdacp$ decay modes, we use the the beam-constrained mass $\mbc$, which is defined as
\begin{eqnarray}
\mbc \equiv \sqrt{{\ebeam}^2/c^4-\left|\vec{p}_{\Lambda_c}\right|^2/c^2},
\label{eq:mbctag}
\end{eqnarray}
where $\vec{p}_{\Lambda_c}$ is the three-momentum of the tagged $\lambdacp$ candidate in the rest frame of the initial $\ee$ collision system. 
To mitigate systematic uncertainties associate with the $\Sigma$ detection, we measure the 
relative BFs of $R_{\sigmodefs{1}}\equiv \BR(\sigmode{1})/\BR(\refmode{1})$ and  $R_{\sigmodefs{2}}\equiv \BR(\sigmode{2})/\BR(\refmode{2})$.

To determine $R_{\sigmodefs{1}}$, the $\mbc$ distributions of the $\sigmode{1}$ and $\refmodefs{1}$ decays, as illustrated in Fig.~\ref{fig:fit Sigma Kp}, are fitted simultaneously for the seven energy points. In the unbinned extended maximum-likelihood fit, the signal shapes are derived from MC simulations convolved with Gaussian functions to account for the  potential difference between data the MC simulations, due to imperfect modeling in MC simulation and the beam-energy spread. The parameters of the Gaussian functions in the signal mode are the same as the reference mode, which are floating in the fit. The combinatorial background is described by an ARGUS function~\cite{Albrecht:1990am}
\begin{equation}
	f^\mathrm{ARGUS} \propto M_{\mathrm{BC}} \sqrt{1-\left(\frac{M_{\mathrm{BC}}\cdot c^2}{E_\mathrm{beam}}\right)^{2}} \times e^{a\left(1-\frac{M_{\mathrm{BC}}\cdot c^2}{E_\mathrm{beam}}\right)^{2}},
	\label{eq:argus}
\end{equation}
where the parameter $a$ is different in the signal and reference modes and the cut-off parameter is fixed to the beam energy $E_\mathrm{beam}$ at each energy point. The observed yield $n_{\sigmodefs{1}}$ of  $\sigmode{1}$ decays is related to  $n_{\refmodefs{1}}$, the yield of $\refmode{1}$ decays, by
\begin{equation}
	n_{\sigmodefs{1}}= R_{\sigmodefs{1}} \cdot  \frac{\varepsilon_{\sigmodefs{1}}}{\varepsilon_{\refmodefs{1}}} \cdot n_{\refmodefs{1}},
\end{equation}
where $\varepsilon_{\sigmodefs{1}}$ ($\varepsilon_{\refmodefs{1}}$) is the detection efficiency of $\sigmode{1}$ ($\refmode{1}$) estimated with the signal MC samples. The relative BF $R_{\sigmodefs{1}}$ is obtained directly in the simultaneous fit as summarized in Table~\ref{tab:fit result}. The fit results are shown in Fig.~\ref{fig:fit Sigma Kp}.

\begin{figure*}[hthp]\centering
    \includegraphics[width=\textwidth]{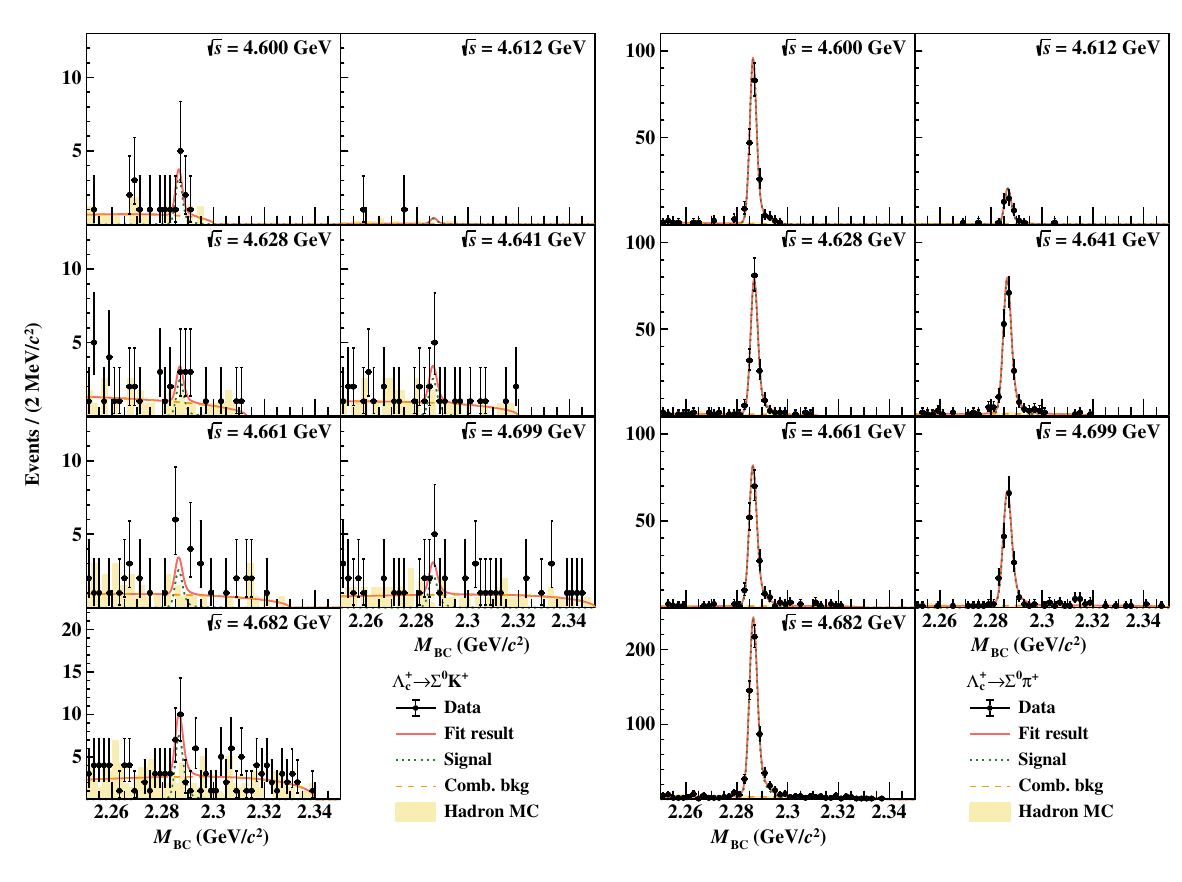}
    \includegraphics[width=0.4\textwidth]{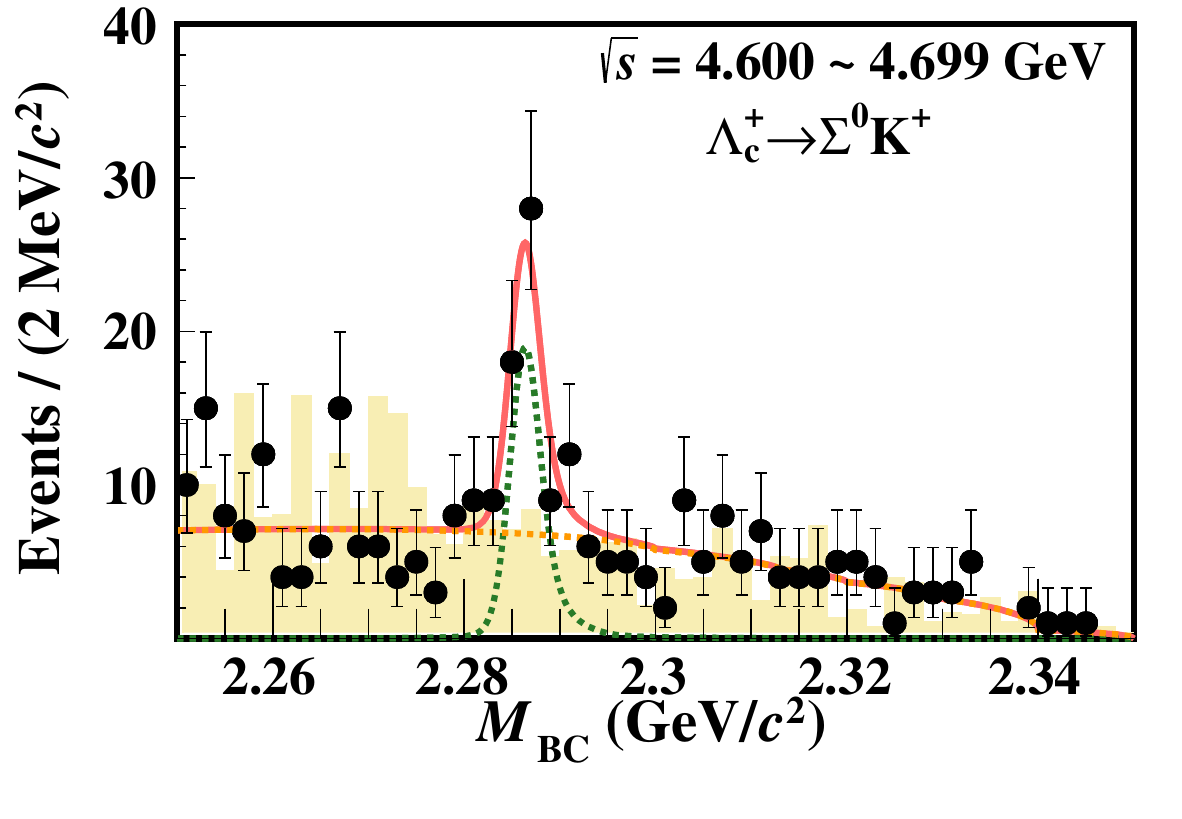}
    \includegraphics[width=0.4\textwidth]{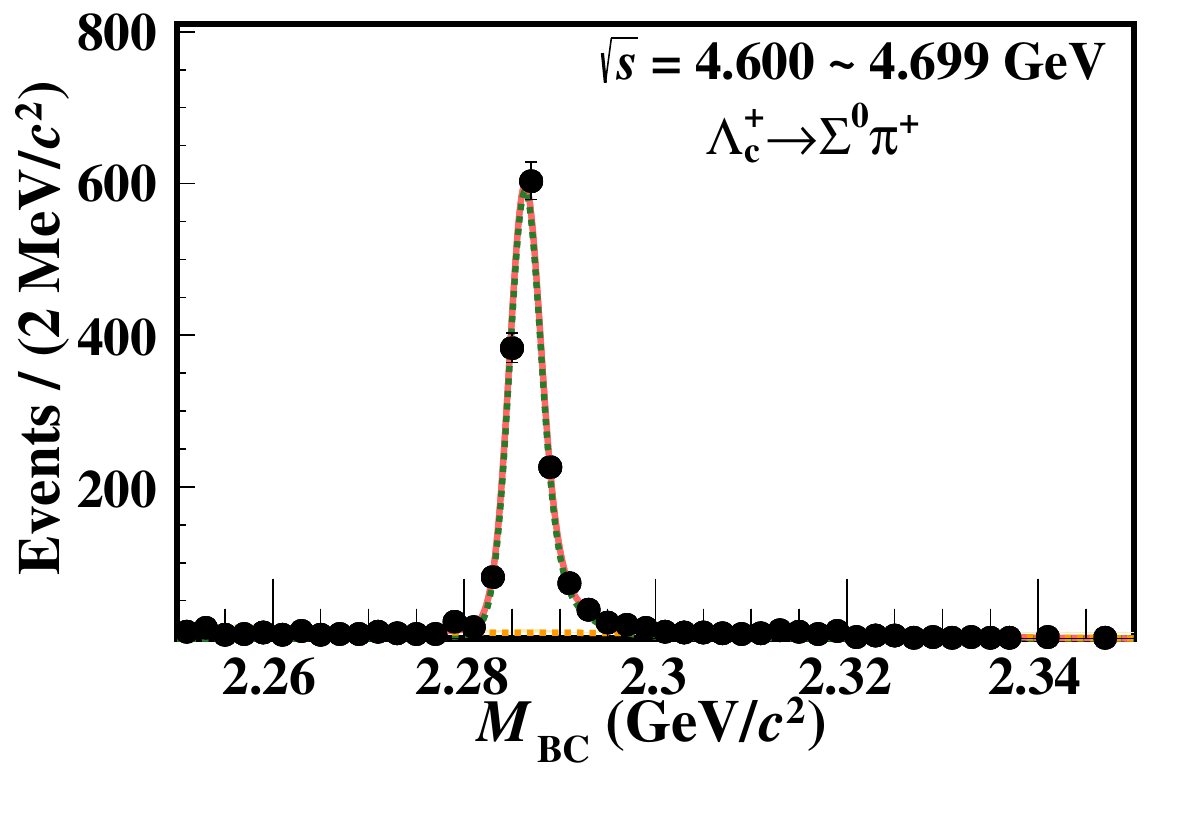}
    \caption{The simultaneous fit results of the $\mbc$ distributions for the $\sigmode{1}$ and $\refmodefs{1}$ candidates at different energy points, where the black points with error bars denote data, the red solid lines denote the fit results, the green dotted lines denote the signal components and the orange dashed lines denote combinatorial backgrounds.}
	\label{fig:fit Sigma Kp}
\end{figure*}

The decay $\bkgmode$ includes many resonant enhancements which may peak in the $\mbc$ spectrum and potentially bias the measurement of  $R_{\sigmodefs{2}}$. A two-dimensional, \emph{i.e.}, $\mbc$ and $M_{p\piz}$, unbinned fit is performed to distinguish the signal $\sigmode{2}$ decay from other contributions. 



In the simultaneous fit to the signal and the reference modes, we adopt two-dimensional signal shapes from MC simulations convolved with two uncorrelated Gaussian functions to determine the yield $n_{\sigmodefs{2}}$ for $\sigmode{2}$ decays, which is related to $n_{\refmodefs{2}}$ by  
\begin{equation}
	n_{\sigmodefs{2}}= R_{\sigmodefs{2}} \cdot \frac{\mathcal{B}_\mathrm{inter}\varepsilon_{\sigmodefs{2}}}{\varepsilon_{\refmodefs{2}}} \cdot n_{\refmodefs{2}},
	\label{eq:KsBF}
\end{equation}
where $\varepsilon_{\sigmodefs{2}}$ ($\varepsilon_{\refmodefs{2}}$) is the detection efficiency of $\sigmode{2}$ ($\refmodefs{2}$) estimated with signal MC samples, and $\mathcal{B}_\mathrm{inter}=\mathcal{B}(\kshort\rightarrow\pi^+\pi^-)$ is taken from the PDG~\cite{Zyla:2020zbs}.
Three types of background components are included: (a) combinatorial backgrounds with no peaking structures in both dimensions, (b) non-$\Sigma^{+}$ background (mainly coming from non-signal $\bkgmode$ decays), and (c) non-$\lambdacp$ background from inclusive $\Sigma^{+}$ production. The combinatorial background component is described by a product of an ARGUS function in the $\mbc$ dimension and a linear function in the $\msigmap$ dimension
\begin{equation}
	f^\mathrm{CombBkg}\propto f^\mathrm{ARGUS} \cdot\left[c_0 + c_1 \msigmap\right],
\end{equation}
where the parameter $a$ of the $f_\mathrm{ARGUS}$ function is different from that in the fit of $\lambdacp\to \Sigma^0 K^+$ and $\Sigma^0 \pi^+$ decays, and $c_0$,  $c_1$ are coefficients of the linear function. 
The approach for modeling the non-$\Sigma^{+}$ background is different between the signal and reference modes. For the reference mode, the shape of non-$\Sigma^{+}$ background is modeled with a product of the $\lambdacp$ signal shape in $\mbc$ and a linear function in $\msigmap$. For the signal mode, the shape of the non-$\Sigma^{+}$ background coming from $\bkgmode$ decays is described by a two-dimensional shape derived from MC simulation in which intermediate states of the process $\bkgmode$  are considered. Other non-$\Sigma^+$ backgrounds from $\lambdacp\to p\eta, p\omega$, \emph{etc.}, are found to be negligible. The shape of non-$\lambdacp$ backgrounds is described by a product of an ARGUS function in $\mbc$ and the $\Sigma^{+}$ signal shape derived from MC simulations in $\msigmap$. The yields of these background components are free parameters in the fit.  When fitting  $\sigmode{2}$, an extra background component is included that accounts for non-$K_S^0$ contamination from the $\refmode{2}$ decays, whose 
shape is fixed according to MC simulations. The yield $n_\mathrm{cont}$ of this background is related to the fitted yield $n_{\refmodefs{2}}$ in the $\refmode{2}$ decay as follows:
\begin{equation}
	n_\mathrm{cont} = \frac{\varepsilon_\mathrm{cont}}{\varepsilon_{\refmodefs{2}}} \cdot n_{\refmodefs{2}}.
\end{equation}
Here $\varepsilon_\mathrm{cont}$ is the contamination rate of $\refmode{2}$ decays into the  $\sigmode{2}$ sample. The relative BF $R_{\sigmodefs{2}}$ is obtained from the simultaneous fit of  $\sigmode{2}$ and $\refmode{2}$ decays, as summarized in Table~\ref{tab:fit result}. The $\mbc$ and $M_{p\pi^0}$ projections of the simultaneous two-dimensional fit of the signal and reference modes are shown in Fig.~\ref{fig:fit Sigma KS} and Fig.~\ref{fig:fit Sigma KS mppi0}.



\begin{table*}[tbp]
	\centering
	\caption{Integrated luminosities taken from Refs.~\cite{BESIII:2015qfd,BESIII:2022ulv}, yields and detection efficiencies for the signal modes $\sigmode{1}$ and $\sigmodefs{2}$, as well as those for the reference modes $\refmode{1}$ and $\refmodefs{2}$ modes at the seven energy points. The uncertainties are  statistical. Also listed are the results for the relative BFs, where the first uncertainties are statistical, and the second are systematic.
	\label{tab:fit result}}
	\bgroup
	\def\arraystretch{1.3}
   \begin{tabular}{ccccccccccc}
	   \hline\hline
	   {$\sqrt{s}$ (GeV)} & $\mathcal{L}_{\mathrm{int}}$ (pb$^{-1}$) &
	   {$n_{\sigmodefs{1}}$} & $n_{\refmodefs{1}}$ &
	   {$10^2\varepsilon_{\sigmodefs{1}}$} & {$10^2\varepsilon_{\refmodefs{1}}$} & \vline & 
	   {$n_{\sigmodefs{2}}$} & {$n_{\refmodefs{2}}$} &	   
	   {$10^2\varepsilon_{\sigmodefs{2}}$} & {$10^2\varepsilon_{\refmodefs{2}}$}\\  
	   \hline
	    4.600 & 566.9 & $5.5\pm1.2$ & $178 \pm 13$  & $9.30\pm 0.04$ & $10.79  \pm 0.05$ & \vline & $6.3\pm1.9$  & $900 \pm 35  $ & $18.85 \pm 0.06$ & $19.92 \pm 0.06$ \\
		4.612 & 103.8 & $1.1\pm0.3$ & $38 \pm 6  $  & $8.46\pm 0.04$ & $10.02 \pm 0.05$  & \vline & $1.1\pm0.3$  & $157 \pm 15  $ & $17.83 \pm 0.06$ & $19.26 \pm 0.06$ \\
		4.628 & 521.5 & $4.8\pm1.0$ & $156 \pm 13$  & $8.31\pm 0.04$ & $9.71  \pm 0.04$  & \vline & $5.3\pm1.6$  & $778 \pm 35  $ & $17.37 \pm 0.06$ & $18.83 \pm 0.06$ \\
		4.641 & 552.4 & $5.4\pm1.2$ & $174 \pm 14$  & $8.36\pm 0.04$ & $9.76 \pm 0.04$   & \vline & $5.2\pm1.6$  & $783 \pm 36  $ & $17.00 \pm 0.06$ & $18.65 \pm 0.06$ \\
		4.661 & 529.6 & $5.3\pm1.1$ & $173 \pm 14$  & $8.20\pm 0.04$ & $9.60  \pm 0.04$  & \vline & $5.2\pm1.6$  & $770 \pm 36  $ & $16.63 \pm 0.06$ & $18.18 \pm 0.06$ \\
		4.682 & 1669.3 & $16.4\pm3.3$& $530 \pm 24$  & $8.22\pm 0.04$ & $9.56 \pm 0.04$   & \vline & $15.9\pm4.8$ & $2395 \pm 63 $ & $16.18 \pm 0.06$ & $17.92 \pm 0.06$ \\
		4.699 & 536.5 & $4.9\pm1.1$ & $159 \pm 13$  & $8.12\pm 0.04$ & $9.47 \pm 0.04$   & \vline & $4.6\pm1.4$  & $694 \pm 34  $ & $15.73 \pm 0.06$ & $17.62 \pm 0.06$ \\

		\hline
			  & \multicolumn{5}{c}{$R_{\sigmodefs{1}} = 0.0361\pm0.0073\pm0.0005$} & \vline &
				\multicolumn{4}{c}{$R_{\sigmodefs{2}} = 0.0106\pm0.0031\pm0.0004$}  \\
	   \hline\hline
   \end{tabular}
   \egroup
\end{table*}


\begin{figure*}[hthp]\centering
    \includegraphics[width=\textwidth]{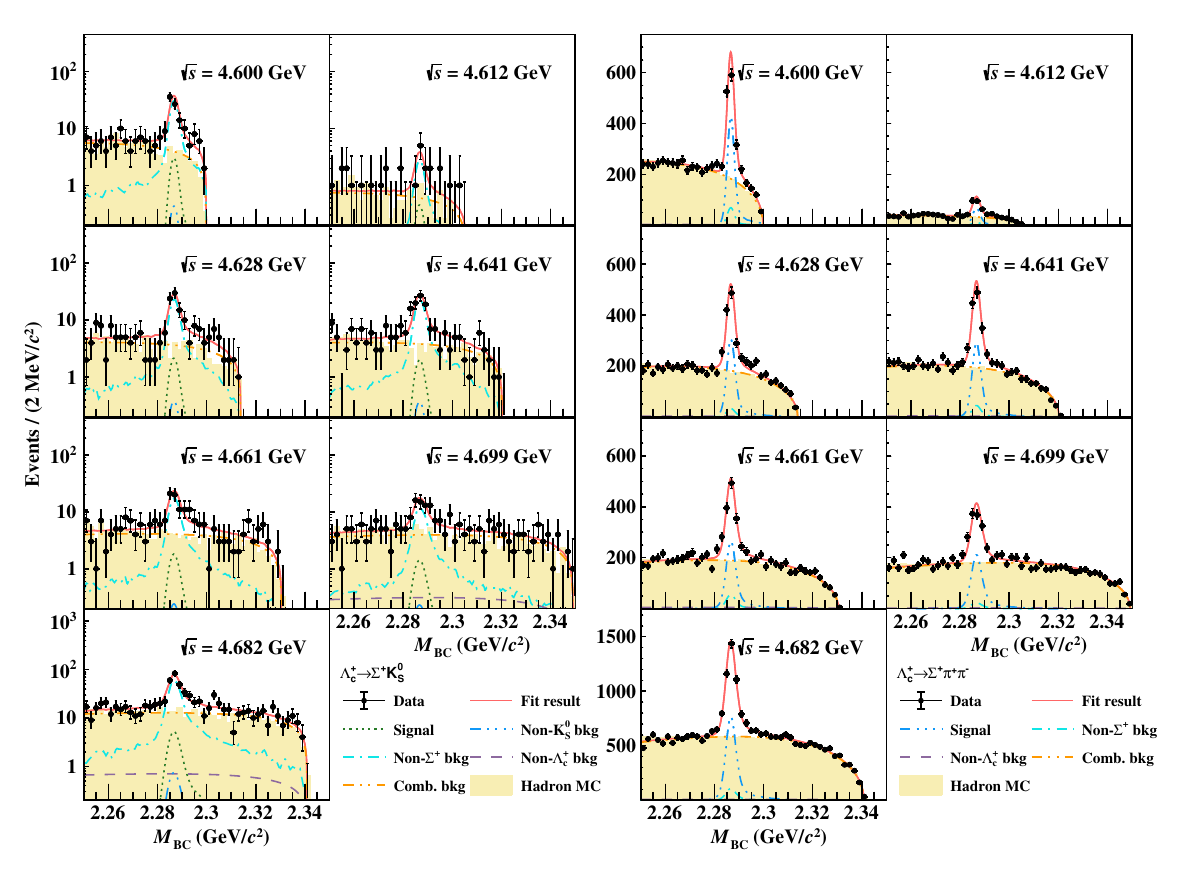}
    \includegraphics[width=0.4\textwidth]{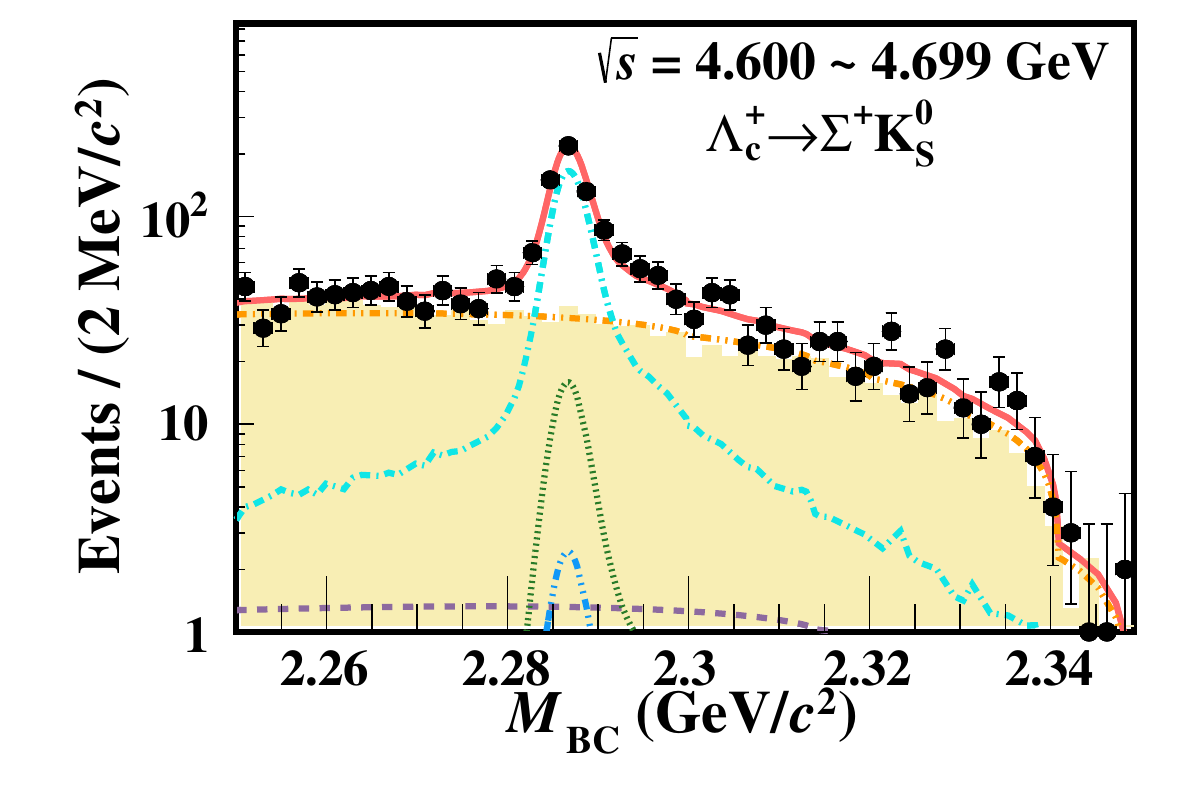}
    \includegraphics[width=0.4\textwidth]{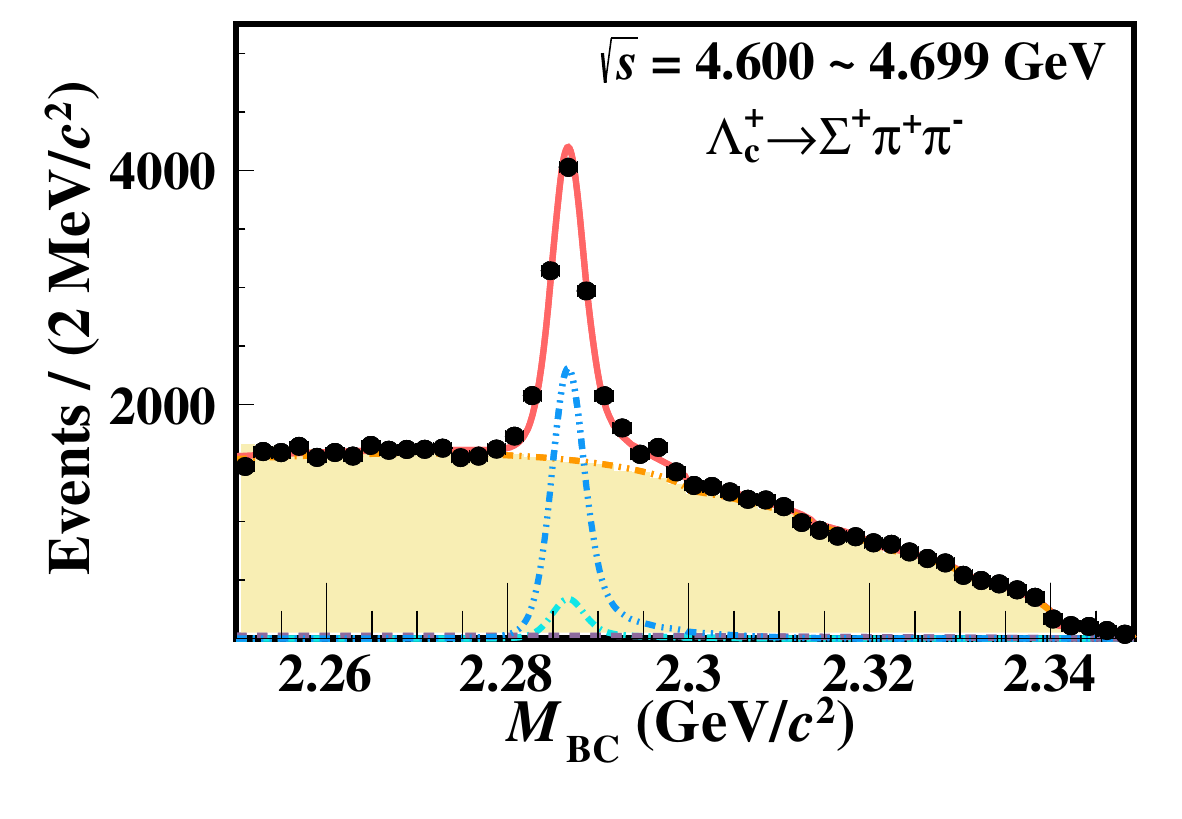}
    \caption{The $\mbc$ projections of the simultaneous two-dimensional fit  of the $\sigmode{2}$ and $\refmodefs{2}$ candidates at different energy points, where the black points with error bars denote data, the red solid lines denote the fit results, and the other colored curves denote the different components. In the left panel, the green dotted lines denote $\sigmode{2}$ signal. In the right side, the blue dashed-dotted lines denote $\refmode{2}$ signal.}
	\label{fig:fit Sigma KS}
\end{figure*}

\begin{figure*}[hthp]\centering
    \includegraphics[width=\textwidth]{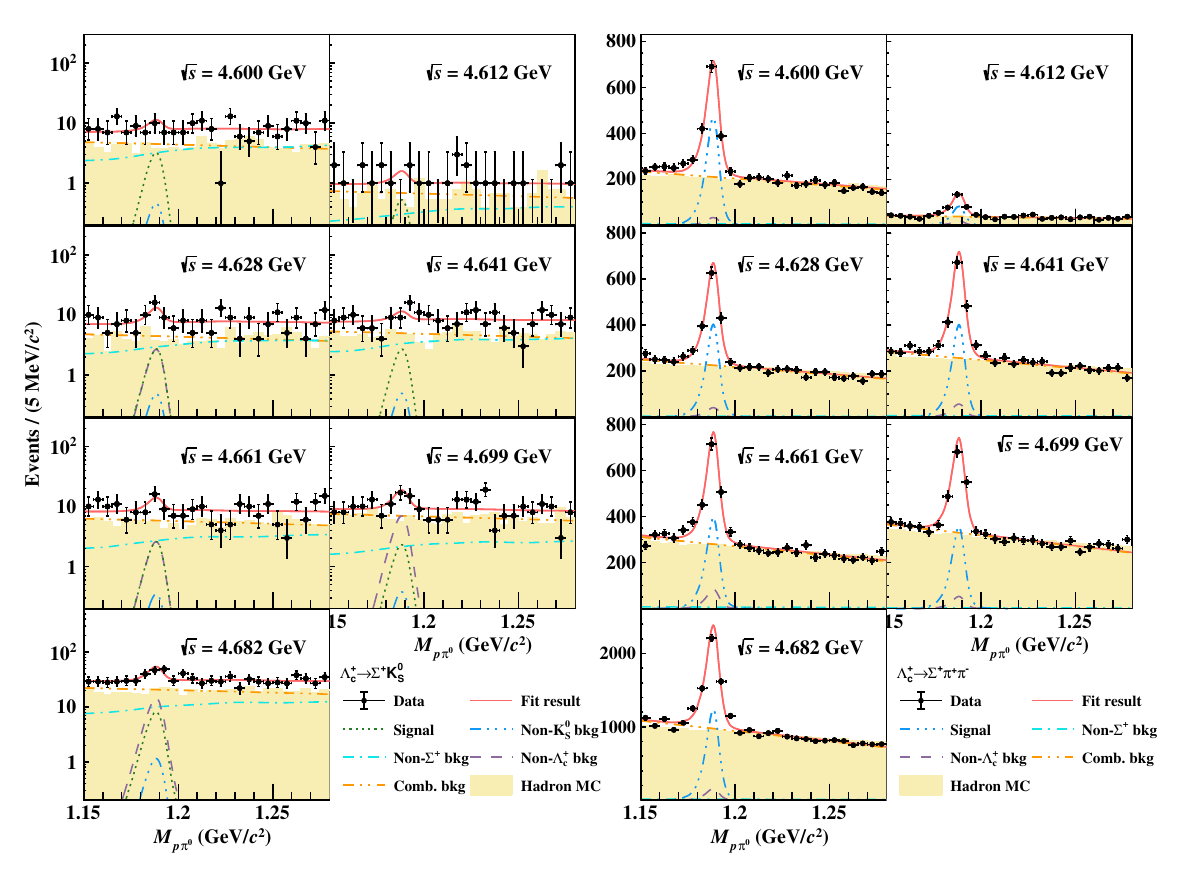}
    \includegraphics[width=0.4\textwidth]{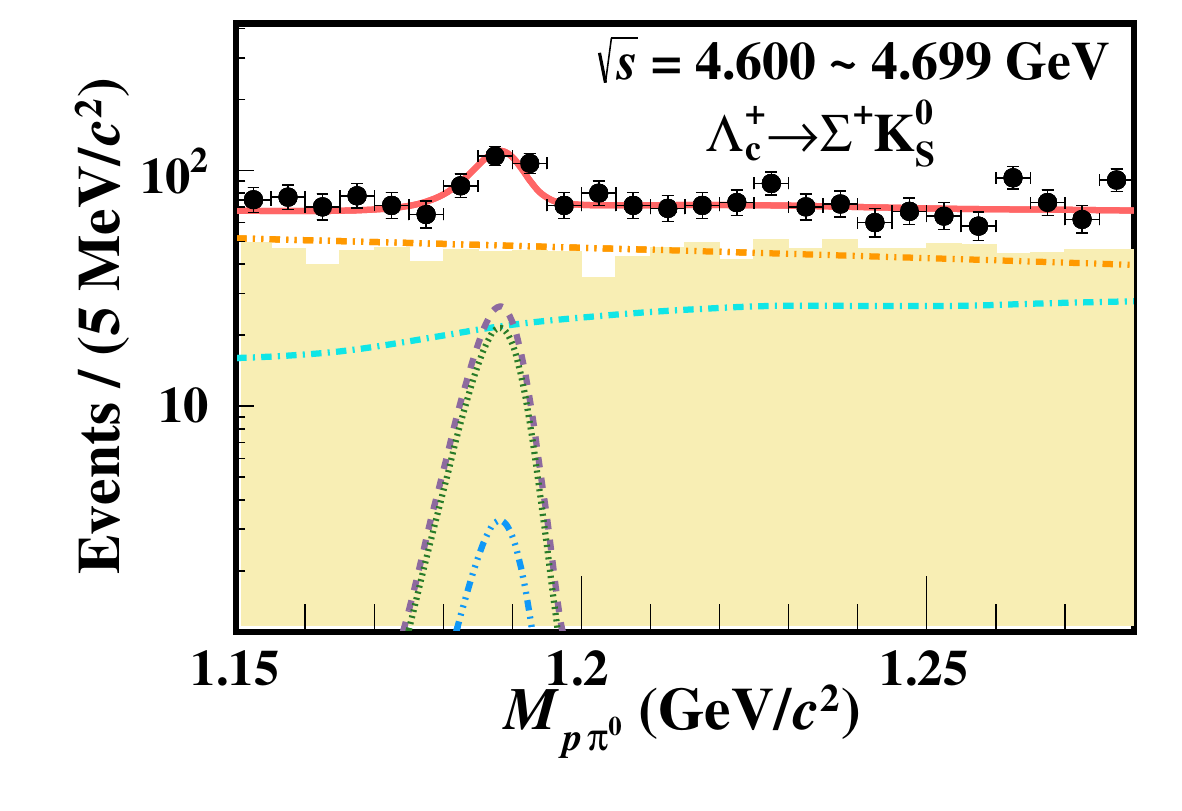}
    \includegraphics[width=0.4\textwidth]{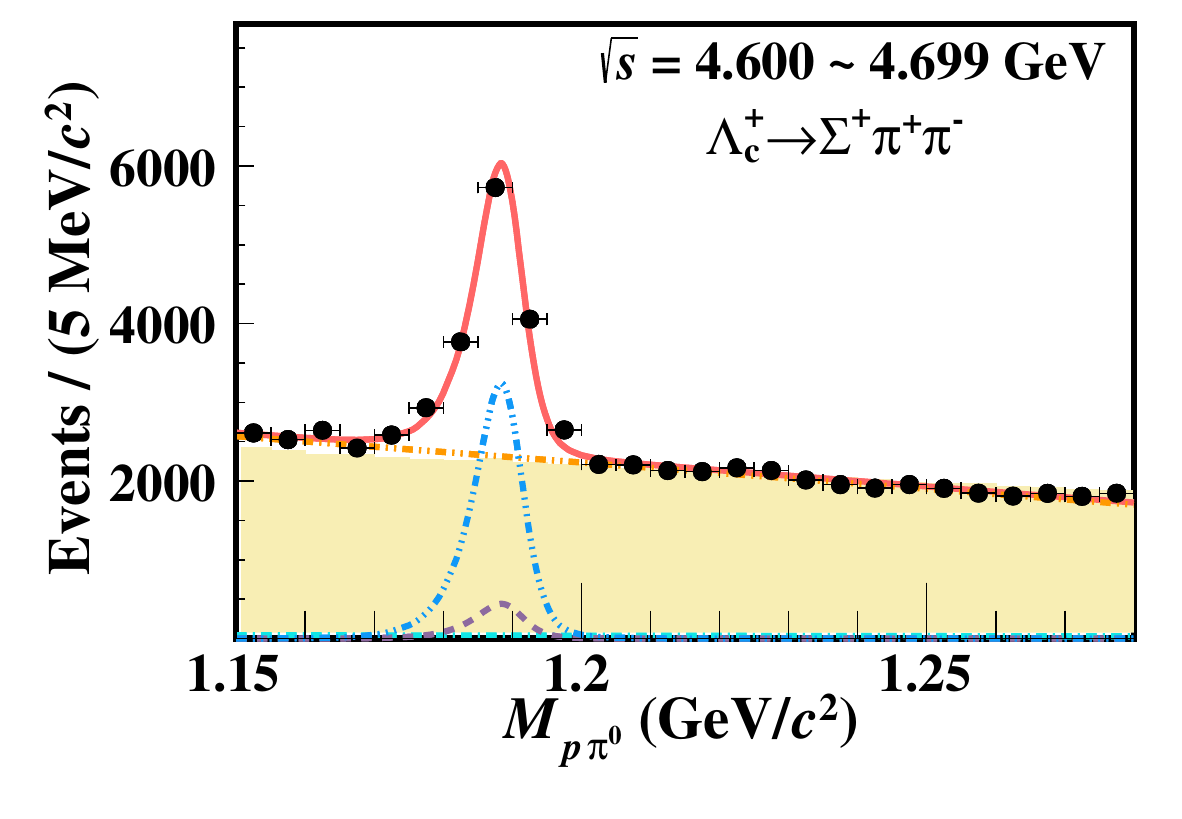}
    \caption{The $M_{p\pi^0}$ projections of the simultaneous two-dimensional fit  of the $\sigmode{2}$ and $\refmodefs{2}$ candidates at different energy points, where the black points with error bars denote data, the red solid lines denote the fit results, and the other colored curves denote the different components. In the left panel, the green dotted lines denote $\sigmode{2}$ signal. In the right side, the blue dashed-dotted lines denote $\refmode{2}$ signal.}
	\label{fig:fit Sigma KS mppi0}
\end{figure*}




\section{Systematic uncertainties}

In the measurements of $R_{\sigmodefs{1}}$ and $R_{\sigmodefs{2}}$, the uncertainties associated with the $\Sigma^0$ and $\Sigma^+$ reconstruction cancel in the ratio. In the $R_{\sigmodefs{1}}$ measurement, the equal number of charged tracks in the signal and reference modes means that any uncertainty in the tracking efficiency also cancels. Similarly, in the $R_{\sigmodefs{2}}$ measurement, there in no uncertainty from PID as the same $\pi^+\pi^-$ PID requirements are  imposed in the $\ks$ reconstruction and for the reference mode $\refmode{2}$. Those uncertainties that do not cancel  are summarized in Table~\ref{tab:syst summary} and discussed below.

\begin{table}[tbp]
   \centering
   \caption{Summary of systematic uncertainties for relative BF measurements of the $\sigmode{1}$ and  $\sigmodefs{2}$ decays. The total systematic uncertainty is the sum in quadrature of the individual components. ``-----'' indicates cases where there is no uncertainty.\label{tab:syst summary}}
   \begin{tabular}{ lcc }
      \hline\hline
	  Source  & $R_{\sigmodefs{1}}$(\%) &  $R_{\sigmodefs{2}}$(\%) \\
      \hline
	  PID &
      0.9      &	-----
               \\
	  Tracking &
      -----      &	1.0
               \\

	  $\kshort$ reconstruction &
      -----        &	1.0
               \\

	  $\kshort$ veto region &
	  -----		   &    0.4
	  		   \\
	  		   
	  $\Lambda$ veto region &
	  -----		   &    0.4
	  		   \\
	  		   
	  $\chi^{2}$ requirement &
      0.7         &		-----
               \\
               
      $\Delta E$ requirement &
      -----			&	0.5
      			\\

	  Signal model &
          0.5   &	2.5
               \\

	  Fitting model &
          0.6    &	1.6
               \\
	  $\BR (\kshort\rightarrow\pi^+\pi^-)$ &
	  -----	   &	0.1
				\\
      \hline
	  Total &
	  	1.4	&	3.4
	  			\\
      \hline
   \end{tabular}
\end{table}

(\romanOne) \emph{PID}. To account for the difference between data and MC simulation, we study a series of control samples of 
$e^{+} e^{-} \rightarrow K^{+} K^{-} \pi^{+} \pi^{-}, K^{+} K^{-} K^{+} K^{-}$, $K^{+} K^{-} \pi^{+} \pi^{-} \pi^{0}, \pi^{+} \pi^{-} \pi^{+} \pi^{-}, \pi^{+} \pi^{-} \pi^{+} \pi^{-} \pi^{0}$ events~\cite{PID_Tracking} to determine the $K^\pm$ and $\pi^\pm$ PID efficiencies. The detection efficiencies for the four $\lambdacp$ decay modes are recalculated after reweighting the corresponding MC samples on an event-by-event basis according to the momentum-dependent efficiency differences between data and MC simulations. The corrected efficiencies are then  input to the simultaneous fits and the resultant changes are taken as the systematic uncertainty. The statistical fluctuations in the control samples also induce uncertainties in the reweighting procedure, which are evaluated  in the different momentum regions and propagated to the final measurement. The average uncertainty is calculated as
		\begin{equation}
			\sigma_{\mathrm{syst.}} = \sqrt{\frac{1}{n_{\mathrm{MC}}}\sum^{n_{\mathrm{MC}}}_{i=1}\sigma_{i}^{2}},
		\end{equation}
where $n_{\mathrm{MC}}$ is the size of the MC samples and $\sigma_{i}$ is the statistical uncertainty of the control samples in the momentum region where the $i$-th event in the MC sample is found. We add these two contributions in quadrature to arrive at the total uncertainty from PID, which is 0.9\% for $R_{\sigmodefs{1}}$.

(\romanTwo) \emph{Tracking}. Using the same control samples to determine the  $\pi^\pm$ tracking efficiencies, we reweight the detection efficiency for $\refmode{2}$ and evaluate the systematic uncertainty in the same way as for the  PID, which results in a contribution of 1.0\% for $R_{\sigmodefs{2}}$.

(\romanThree) \emph{$\kshort$ reconstruction}. The $\kshort$ reconstruction efficiencies are studied by using control samples of $J/\psi \rightarrow K^{*}(892)^{\pm} K^{\mp}$, $K^{*}(892)^{\pm} \rightarrow K_{S}^{0} \pi^{\pm}$and $J/\psi \rightarrow \phi K_{S}^{0} K^{\mp} \pi^{\pm}$ decays. The systematic uncertainty is again evaluated in the same way as for the PID, which gives a contribution of 1.0\% for $R_{\sigmodefs{2}}$.

(\romanFour) \emph{$\kshort(\Lambda)$ veto region}. For the $\refmode{2}$ decay mode,  the $\kshort(\Lambda)$ veto may have a different efficiency in data and MC simulation.  To bound any such bias, we remove the $\kshort(\Lambda)$ veto and reevaluate the relative BFs. The relative difference between the reevaluated result and the nominal result is taken as the systematic uncertainty, which is 0.4\%(0.4\%) for $R_{\sigmodefs{2}}$.
 
(\romanFive) \emph{Joint fit-quality $\chi^2_\mathrm{total}$ requirement}. Joint fit-quality $\chi^2_\mathrm{total}$ requirements are imposed on the $\sigmode{1}$ and $\refmodefs{1}$ decay modes. To evaluate the effect of any difference between data and MC simulation, we fit to the $\chi^2_\mathrm{total}$ distribution in data with the shapes derived from MC simulations convolved with a floating Gaussian function.
The detection efficiency is remeasured after resampling the corresponding $\chi^2_\mathrm{total}$ variable in MC samples according to the fitted Gaussian function, and $R_{\sigmodefs{1}}$ is updated accordingly.
The Gaussian parameters are randomly varied within the fitted uncertainties, and the corresponding value of $R_{\sigmodefs{1}}$ is recorded.  The mean $\mu_{\chi^2}$ and the standard deviation $\sigma_{\chi^2}$ of the $R_{\sigmodefs{1}}$ distribution is measured and $\sqrt{\mu_{\chi^2}^2+\sigma_{\chi^2}^2}/R_{\sigmodefs{1}}$ is assigned as the the corresponding systematic uncertainty, which is 0.7\%.

(\romanSix) \emph{Energy difference $\dE$ requirement}. The energy difference $\dE$ requirement is used in the measurement of $R_{\sigmodefs{2}}$. Following a similar procedure  as for the  $\chi^2$ uncertainty, we first obtain the difference of $\dE$ distributions in data and MC simulations, and then study the distribution of the corresponding $R_{\sigmodefs{2}}$ values with updated efficiencies, which leads to the assignment of a systematic uncertainty of 0.5\% for $R_{\sigmodefs{2}}$.

(\romanSeven) \emph{Signal model}. We use the phase-space model for simulating the $\sigmode{1}$ and $\sigmodefs{2}$ decay modes to obtain our central values, since the decay-asymmetry parameters are not known. As a systematic check, we test several extreme cases of the decay-asymmetry parameters and recalculate their detection efficiencies based on the corresponding signal MC samples. The resultant changes on the relative BFs are taken as the systematic uncertainties, which are 0.5\% and 2.5\% for $R_{\sigmodefs{1}}$ and $R_{\sigmodefs{2}}$, respectively.

(\romanEight) \emph{Fitting model}. The fitting procedure has uncertainties arising from both the signal and background shapes. For the signal shapes, the convolved Gaussian functions are varied by $\pm 1 \sigma$ around their baseline values. For the $\sigmode{2}$ mode, we also consider the effect of the correlation between the convolved Gaussian functions in $\mbc$ and $M_{p\piz}$ by including a correlation coefficient determined from MC and verify that this leads to a negligible bias. 
 For the background shapes, the parameter $\ebeam$ in the ARGUS function of Eq.~\eqref{eq:argus} is randomly varied by $\pm0.15$ MeV and the shape of the $\bkgmode$ background component is replaced by a shape from an alternative partial wave analysis, in which more intermediate states are considered than those in the nominal signal MC simulation.
To take into account these effects, 5000 pseudo data sets are sampled  according to the bootstrap method~\cite{chernick2011bootstrap}. 
For each pseudo data set,
the fitting models are varied randomly.
Pull distributions are inspected from these pseudo data sets, and the small biases of  0.6\% and 1.6\% are assigned as the systematic uncertainties for $R_{\sigmodefs{1}}$ and $R_{\sigmodefs{2}}$, respectively.


(\romanNine) \emph{$\BR (\kshort\rightarrow\pi^+\pi^-)$}. The uncertainty on the BF $\BR(\kshort\rightarrow\pi^+\pi^-)$~\cite{Zyla:2020zbs} is propagated to give a systematic uncertainty of 0.1\%  in  $R_{\sigmodefs{2}}$.

\section{\boldmath Summary}
In summary, 
based on 4.4 $\ifb$ of $\ee$ annihilation data collected in the energy region between 4.6 $\gev$ and 4.7 $\gev$ with the BESIII detector at
BEPCII, we report the measurements of BFs of the two singly Cabibbo-suppressed decay modes $\Lambda_c^+ \rightarrow \Sigma^0 K^+$ and $\Sigma^+ \kshort$ relative to the reference modes $\refmode{1}$ and  $\refmodefs{2}$, respectively. The BF of $\Lambda_c^+ \rightarrow \Sigma^0 K^+$ relative to $\Lambda_c^+ \rightarrow \Sigma^0 \pi^+$ is measured to be $0.0361 \pm 0.0073(\mathrm{stat.}) \pm 0.0005(\mathrm{syst.})$, while the BF of $\Lambda_c^+ \rightarrow \Sigma^+ \kshort$  relative to $\Lambda_{c}^{+} \rightarrow \Sigma^+ \pip \pim$ is measured to be $0.0106 \pm 0.0031(\mathrm{stat.}) \pm 0.0004(\mathrm{syst.})$. 
Taking the world-average BFs $\BR(\refmode{1})=(1.29\pm0.07)\%$ and $\BR(\refmode{2})=(4.50\pm0.25)\%$ from the PDG~\cite{Zyla:2020zbs},
yields the the absolute BF $\BR(\sigmode{1}) = (4.7\pm 0.9(\mathrm{stat.})\pm 0.1(\mathrm{syst.}) \pm 0.3(\mathrm{ref.}))\times10^{-4}$,
and $\BR(\sigmode{2}) = (4.8\pm 1.4(\mathrm{stat.})\pm 0.2(\mathrm{syst.}) \pm 0.3(\mathrm{ref.}))\times10^{-4}$.
This is the first measurement of the $\sigmode{2}$ branching fraction. The $\sigmode{1}$ BF is measured with a comparable precision to the combined result from the  Belle~\cite{Abe:2001mb} and BaBar~\cite{Aubert:2006wm} collaborations. 
The ratio $\BR(\sigmode{1})/\BR(\sigmode{2})$ is determined to be $0.98 \pm 0.35(\mathrm{stat.})\pm 0.04(\mathrm{syst.})\pm 0.08(\mathrm{ref.})$, which is consistent with the predictions in Refs.~\cite{Zou:2020, Geng:2019xbo} under $SU(3)_F$ flavor symmetry and disfavors the prediction in Ref.~\cite{he2021global}. The prediction for $\BR(\sigmode{2})$ in Ref.~\cite{he2021global} differs from our result by 2.5$\sigma$, indicating a reassessment of the IRA method may be needed. Though our work is generally consistent with the predictions in Refs.~\cite{Uppal:1994pt, Zou:2020, Zhao:2020, Geng:2019xbo} within $1\sim2\sigma$, these theoretical predictions generally overestimate the BFs.
The systematic uncertainties of our results are smaller than those of  from the Belle and BaBar collaborations,  but  our precision is limited by the relatively low sample size. With the additional data which are foreseen to be collected near the $\lamcplamcm$ threshold in the coming years~\cite{BESIII:2020nme}, we expect our measurements to improve in precision, and shed more light on the topic of charmed baryon decays.

\emph{
Note: After the publication of our measurement, we were contacted by the authors of Ref.~\cite{he2021global} that there was a numerical mistake in their calculation of $\BR(\sigmode{2})$. The erratum can be found in Ref.~\cite{ErratumHuang:2021aqu}. The result has been changed to $\BR(\sigmode{2})=(6.3\pm2.5)\%$, which is consistent with our measurement within $1\sigma$.
}

\acknowledgments
The BESIII collaboration thanks the staff of BEPCII and the IHEP computing center for their strong support. This work is supported in part by National Key R\&D Program of China under Contracts Nos. 2020YFA0406400, 2020YFA0406300; National Natural Science Foundation of China (NSFC) under Contracts Nos. 11635010, 11675275, 11735014, 11822506, 11835012, 11935015, 11935016, 11935018, 11961141012, 12022510, 12025502, 12035009, 12035013, 12175321, 12192260, 12192261, 12192262, 12192263, 12192264, 12192265; the Chinese Academy of Sciences (CAS) Large-Scale Scientific Facility Program; Joint Large-Scale Scientific Facility Funds of the NSFC and CAS under Contract No. U1832207, U1932101; State Key Laboratory of Nuclear Physics and Technology, PKU under Grant No. NPT2020KFY04; CAS Key Research Program of Frontier Sciences under Contract No. QYZDJ-SSW-SLH040; 100 Talents Program of CAS; the Fundamental Research Funds for the Central Universities; INPAC and Shanghai Key Laboratory for Particle Physics and Cosmology; ERC under Contract No. 758462; European Union's Horizon 2020 research and innovation programme under Marie Sklodowska-Curie grant agreement under Contract No. 894790; German Research Foundation DFG under Contracts Nos. 443159800, Collaborative Research Center CRC 1044, GRK 2149; Istituto Nazionale di Fisica Nucleare, Italy; Ministry of Development of Turkey under Contract No. DPT2006K-120470; National Science and Technology fund; National Science Research and Innovation Fund (NSRF) via the Program Management Unit for Human Resources \& Institutional Development, Research and Innovation under Contract No. B16F640076; STFC (United Kingdom); Suranaree University of Technology (SUT), Thailand Science Research and Innovation (TSRI), and National Science Research and Innovation Fund (NSRF) under Contract No. 160355; The Royal Society, UK under Contracts Nos. DH140054, DH160214; The Swedish Research Council; U. S. Department of Energy under Contract No. DE-FG02-05ER41374.


\bibliographystyle{apsrev4-2}
\bibliography{mybib}

\end{document}